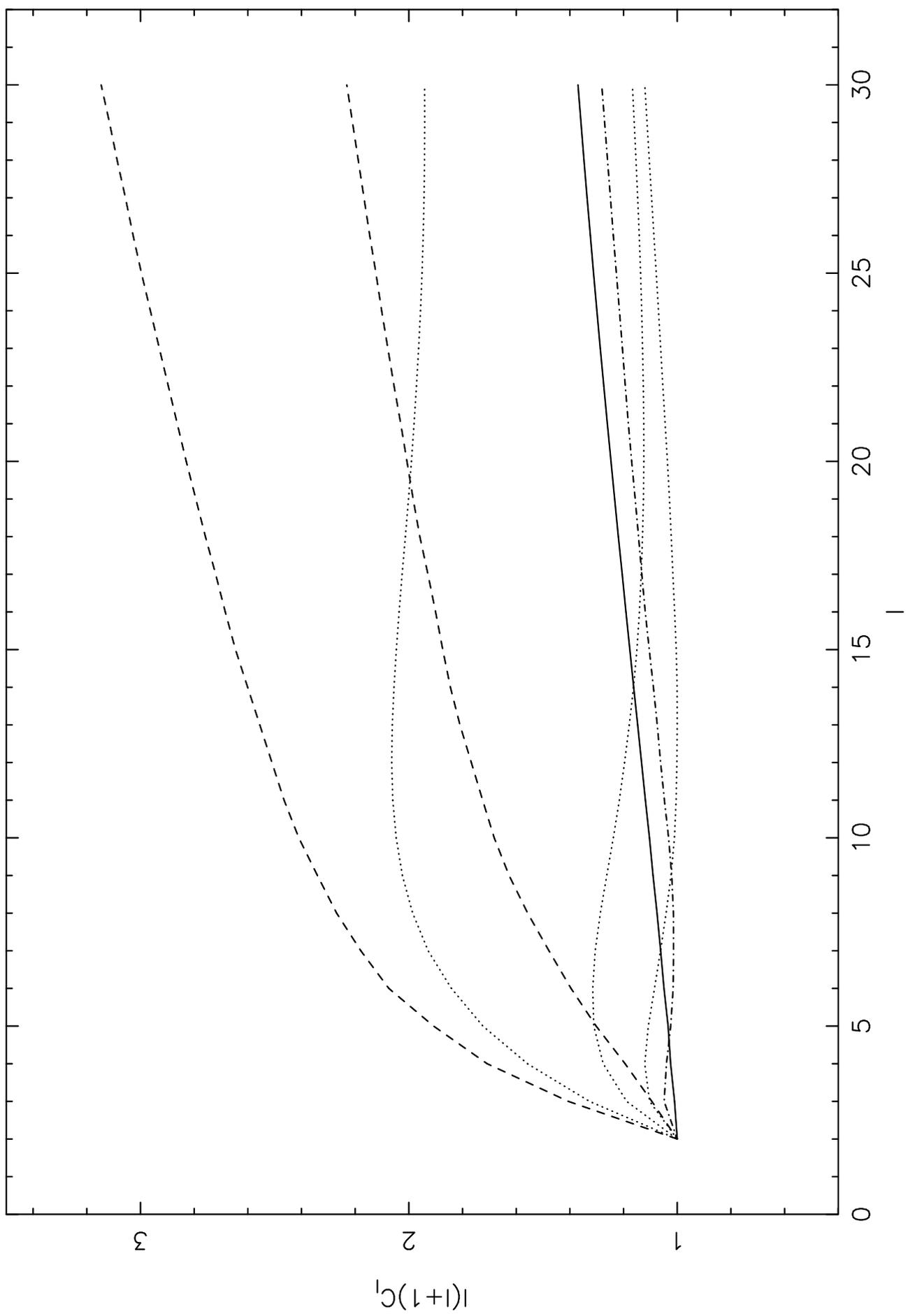

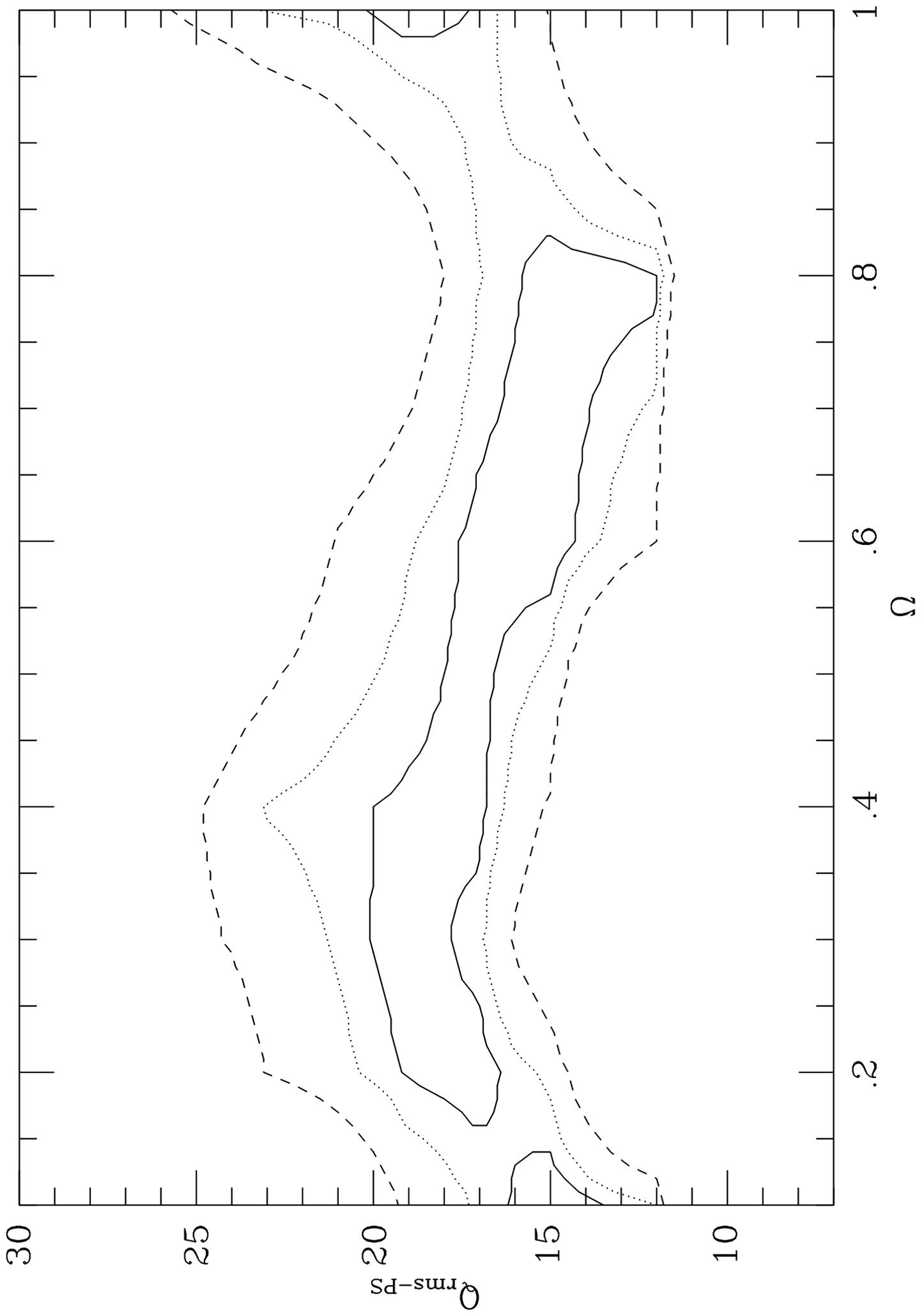

# Ω from the *COBE*-DMR anisotropy maps

L. Cayón[1,2], E. Martínez-González[3], J.L. Sanz[3], N. Sugiyama[2,4] and S. Torres[5]
[1] *Lawrence Berkeley Laboratory, Berkeley, CA, USA.*
[2] *Center for Particle Astrophysics, Berkeley, CA, USA.*
[3] *Instituto de Física de Cantabria, Consejo Superior de Investigaciones Científicas–Universidad de Cantabria, Santander, Spain.*
[4] *Department of Physics, Faculty of Science, University of Tokyo, Tokyo, Japan.*
[5] *Observatorio Astronómico Universidad Nacional and Centro Internacional de Física, Bogotá, Colombia.*

**ABSTRACT**
We have made a likelihood statistical analysis of the angular correlations in the *COBE*-DMR two-year sky maps by Monte Carlo simulation of the temperature fluctuations. We assume an open universe and consider as primordial power spectrum the Harrison-Zeldovich one, $P(k) = Ak$. We find that the flatness of the universe is not implied by the data. The quadrupole normalization amplitude, $Q_{rms-PS}$, is related to the density parameter, $\Omega$, by $Q_{rms-PS} = 10.67 + 55.81\Omega - 128.59\Omega^2 + 81.26\Omega^3$ $\mu$K. We have determined the p.d.f. of $\Omega$ due to cosmic plus sampling (i.e. $20°$ galactic cut) variance which generically shows a bimodal shape. The uncertainty as given by the r.m.s. is $\approx 0.35$, therefore to better constrain $\Omega$ experiments sensitive to higher multipoles ($l > 20$) should be considered.

## 1 INTRODUCTION

Rapid growth of the number of cosmological observations reveals the week points of standard $\Omega = 1$ CDM models. These models are unable to simultaneously account for the following observations: galaxy and cluster autocorrelation functions, large scale streaming motions and the normalization amplitude of the Cosmic Microwave Background (CMB) spectrum. A low-density ($\Omega < 1$) universe is favoured by many of the observations (see, e.g., Coles & Ellis 1994). Other possibilities are the flat universe with a cosmological constant, tilted models and mixed hot-cold dark matter models. Flat and low density models present quite different shapes of the CMB anisotropies due to geometric effects (Sugiyama & Silk 1994). We can use the angular correlation of CMB maps to test open, flat and low density flat models. By assuming a Harrison-Zeldovich (HZ) spectrum, Bunn & Sugiyama (1995) constrained the value of the cosmological constant from *COBE*-DMR maps.

Open universes have not been studied in detail because the precise shape of the density power spectrum beyond the curvature scale in these universes has not been determined definitively. The pioneering calculation of Wilson (1983) assumed a power law in wavenumber $k$ which is defined as $k^2 = q^2 - 1$ where $q$ is the eigenvalue of the Laplacian, while Kamionkowski and Spergel (1994) considered power laws in spatial volume and in scalar spatial Laplacian eigenvalue. Recently, the power spectrum has been calculated by assuming inflation at the early epoch of the universe (Lyth & Stwart 1990; Ratra & Peebles 1994). This model has been confronted with some observations which do not include all of the *COBE*-DMR experimental features (Kamionkowski et al. 1994; Sugiyama and Silk 1994) but Gorski et al. (1995), incorporating such features, conclude that the *COBE*-DMR two-year data do not provide sufficient statistical power to prefer a particular value of $\Omega$. However, further modification of the power spectrum might be necessary because of the bubble nucleation in the early epoch which allows the existence of the open universe together with inflation (Bucher, Goldhaber & Turrok 1995; Lyth & Woszczyna 1995; Yamamoto, Sasaki & Tanaka 1995).

In this paper, we study the angular correlation of CMB anisotropies with the simplest possible shape of the initial power spectrum, i.e., the HZ spectrum $P(k) \propto k$ where $k$ is the wave number. CMB large scale anisotropies in a flat universe are produced by density fluctuations on the last scattering surface (Sachs & Wolfe 1967). At angular scales greater than $(2\Omega^{1/2})°$ a generalized formula for curved spaces can be obtained (Anile and Motta 1976). In this case the anisotropies can be expressed in terms of gravitational potential fluctuations at recombination and an integrated effect of the time-varying gravitational potential along the photon trajectory that depends on curvature (Abbott and Schaefer 1986, Traschen and Eardley 1986, Gouda et al. 1991). In this work we have calculated CMB anisotropies by numerically integrating the generalized formulas for curved spaces (Sugiyama & Silk 1994) as shown in Fig. 1. Taking a multipole expansion for the temperature fluctuations

**Figure 1.** The angular power spectrum for various $\Omega$ values: 1 (solid), 0.8, 0.6 (upper and lower dashed lines), 0.4 (dash-dotted), 0.3, 0.2, 0.1 (lower, middle and upper dotted lines. The Harrison-Zeldovich primordial spectrum was assumed.

$\Delta T/T = \Sigma_{l,m} a_{l,m} Y_{lm}$, only multipoles up to 30 are sensitive to the $COBE$-DMR angular resolution.

Analysis of the first year $COBE$-DMR maps using angular correlations resulted in a spectral index, $n$ of $1.1 \pm 0.5$ (assuming a power law spectrum $P(k) = Ak^n$), and a rms-quadrupole-normalized amplitude, $Q_{rms-PS}$, of $16 \pm 4\mu K$ (Smoot et al.1992). An analysis based on the 'genus', or total contour curvature of anisotropy spots, found $n = 1.2 \pm 0.3$ (Torres 1994a, 1994b). This result was obtained by fitting the coherence angle of the $COBE$-DMR temperature maps for a fixed $Q_{rms-PS}$ of $16\mu K$, assuming a power law primordial density fluctuation spectrum. By allowing variations of the $Q_{rms-PS}$ ($\propto A^{1/2}$) parameter, a relation between $Q_{rms-PS}$ and $n$ was also found (Smoot et al. 1994, Torres et al. 1995). Analysis of the first two years $COBE$-DMR maps shows that the Harrison-Zeldovich spectrum remains consistent with the spectral index predicted by the data (Bennett et al. 1994, Gorski et al. 1994). In all the previous analyses a flat universe with $\Omega = 1$ was assumed.

Even in an ideal noise-less experiment the measured parameters will have an uncertainty from cosmic variance. We have used Monte Carlo simulations to study the extent to which cosmic and sampling variance obscures the information about $\Omega$ which can be extracted from a correlation analysis.

In section 2 we present the $COBE$-DMR Monte Carlo simulation methodology while in section 3, we obtain the minimum range of $\Omega$ implied by cosmic and sampling variance. In section 4 we apply the correlation technique to the $COBE$-DMR data for low-$\Omega$ models and in section 5 summarize and discuss the results.

## 2  MONTE CARLO METHODOLOGY

Taking into account instrument noise, non-uniform sky coverage, galactic cut, smearing, pixellization scheme, and the DMR beam characteristics we simulated a set of cosmological models defined by $\Omega$, $Q_{rms-PS}$, and HZ primordial spectrum. We generated 3200 simulated CMB sky maps using a harmonic expansion of the temperature for each of the 6,144 DMR pixels:

$$\Delta T(\theta, \phi) = \sum_{l=2}^{} \sum_{m=0}^{l} k_m [b_{l,m} \cos(m\phi) + b_{l,-m} \sin(m\phi)] N_l^m W_l P_l^m(\cos\theta), \quad (1)$$

where

$$N_l^m = \left[\frac{(2l+1)(l-m)!}{4\pi(l+m)!}\right]^{1/2}.$$

$k_m = \sqrt{2}$ for $m \neq 0$ and $k_0 = 1$. $P_l^m(\cos\theta)$ are the associated Legendre polynomials. The coefficients $b_{l,m}$ are real stochastic, Gaussianly distributed variables with zero mean and model dependent variance $\langle b_{l,m}^2 \rangle$ that are easily obtained from the multipole coefficients $C_l$ ($\langle b_{l,m}^2 \rangle = C_l/4\pi$) given by Sugiyama & Silk (1994) in their Figure 1a. The weights, $W_\ell$, for DMR given by Wright et al. (1994) are used and the series is truncated at $l = 30$. A small beam smearing correction is applied which accounts for the spacecraft motion during the 1/2 second integration time. For each realization two maps are generated by adding to the cosmic signal the noise corresponding to the channel combination (A+B)/2 of $COBE$-DMR 53 GHz and 90 GHz. The noise is determined by instrument sensitivity and the number of observations per pixel. The angular cross-correlation of the 53 GHz and 90 GHz maps for $|b| > 20°$ is calculated and binned in 36, 5° bins in the manner of the $COBE$-DMR data (Bennett et al. 1994). A combination of the two indepedent frequency maps helps reducing the introduction of systematic effects.

The standard likelihood statistic L associated to the cross-corelation is ($n = 36$)

$$L = \frac{1}{(2\pi)^{n/2}(det M)^{1/2}} e^{-\frac{1}{2}\chi^2}, \quad (2)$$

$$\chi^2 = \sum_{i=1}^{36} \sum_{j=1}^{36} (\langle C_i \rangle - C_i^{COBE}) M_{ij}^{-1} (\langle C_j \rangle - C_j^{COBE}). \quad (3)$$

$\langle C_i \rangle$ is the average of the cross-correlation for the 3200 realizations at bin $i$. $C_i^{COBE}$ is the cross-correlation for the $COBE$-DMR maps at angular scale $i$ (note that $C_i$ must not be confused with the multipole coefficients $C_l$). $M_{ij}$ is the covariance matrix calculated with the Monte Carlo realizations:

$$M_{ij} = \frac{1}{N_{realiz.}} \sum_{k=1}^{N_{realiz.}} (C_i^k - \langle C_i \rangle)(C_j^k - \langle C_j \rangle). \quad (4)$$

The likelihood method is preferable to the $\chi^2$ one based on the following test: we have used as 'data' 400 realizations from each of the following $\Omega - Q_{rms-PS}(\mu K)$ pair of values: $0.1 - 15, 0.2 - 17, 0.3 - 19, 0.4 - 17, 0.6 - 15, 0.8 - 15, 1 - 19$ (these are the models with maximum L for fixed $\Omega$, see section 4). For each $\Omega$ we get the distribution of Q-values for which L and $\chi^2$ are maximum and minimum, respectively, by comparing each one of the 400 'data' realizations with 3200 realizations of the models with $Q_{rms-PS} = 7, 12, 15, 17, 19, 21, 23, 25, 27$. The mean value of the L method is within 0.5 units of the true value with a dispersion of $\sigma_Q \approx 1.5$ for all $\Omega$ values. However the mean value of the $\chi^2$ method systematically overestimates the true $Q_{rms-PS}$ value in $\approx 3$ units and with a significantly bigger dispersion. So, the likelihood method accurately predicts the $Q_{rms-PS}$ value with a relatively small uncertainty whereas the $\chi^2$ has systematic errors. This result could be explained by the fact that the number of bins (i.e. 36) begins to be high enough for the likelihood to be unbiased.

## 3 UNCERTAINTY FROM COSMIC PLUS SAMPLING VARIANCE

Assuming a noise-less radiometer, we now study the cosmic variance associated to the cosmological parameters derived from the correlation analysis. The *COBE*-DMR radiometer noise is taken into account in the next section when applying the method to the real data.

The non-ergodic character of the CMB temperature random field on the cosmic sphere limits the accuracy in determining the statistical properties of the field from a single realization (i.e. averages over the celestial sphere do not coincide with statistical averages over the ensemble of realizations). Even in the ideal case when the cosmic background temperature is measured in every part of the celestial sphere with no galactic cut and with negligible noise we would only be able to measure the expected values of the field within certain errors, the cosmic variance. Uncertainties in the two-point correlation function and in other higher moments due to cosmic variance have been calculated (Scaramella and Vittorio 1990, Cayón, Martínez-González and Sanz 1991, White, Krauss and Silk 1993, Srednicki 1993, Gutiérrez de la Cruz et al. 1994). Uncertainty in the determination of the spectral index $n$ for a flat universe has been also estimated in the genus analysis (Torres et al. 1995).

We calculate the cosmic plus sampling variance uncertainty when analysing the cross-correlation of two hypothetical sky maps with the characteristics of the *COBE*-DMR experiment (i.e. 7° FWHM beam) assuming a noise-less radiometer and $|b| > 20°$ galactic cut. Here we have two free parameters, $\Omega$ and $Q_{rms-PS}$, and seven Monte Carlo data sets were generated with 3200 realizations per set, one for each $\Omega$ value $-$ 0.1, 0.2, 0.3, 0.4, 0.6, 0.8,and 1.0 $-$, with fixed $Q_{rms-PS}$ since the results are not too sensitive to changes in this parameter over a realistic range of values. To estimate the uncertainty in $\Omega$ due to cosmic variance we computed the likelihood L for each of 400 realizations from a previously selected model. As a first case $\Omega$ and $Q_{rms-PS}$ were taken to be 0.4 and 17 $\mu$K (this is the normalization for which L is maximum for the model $\Omega = 0.4$, as obtained in the next section). $C_i^{COBE}$ in the likelihood definition was replaced by $C_i^k$, the value of the cross-correlation of the $k$-th realization in bin $i$. For each realization the value of $\Omega$ which maximized L ($\Omega_m$) was found. The resulting distribution of $\Omega_m$ is: 0, 11, 11, 44, 0, 0, 34% probabilities for $\Omega = 0.1, 0.2, 0.3, 0.4, 0.6, 0.8, 1$. Therefore, the flat model is indistinguishable from the 'true' 0.4 model and only models with $\Omega = 0.1, 0.6, 0.8$ are clearly distinguished. This counter-intuitive result can be accounted for by the similarities in the angular power spectrum shown by the $\Omega = 0.4$ and 1 models and the large differences between $\Omega = 0.4$ and 0.8 models (see Fig. 1). In general, for other values of the 'true' $\Omega$ a similar bimodal distribution appears.

**Figure 2.** The likelihood function L (normalized to 1 at the maximum value) as a function of $\Omega$ and $Q_{rms-PS}$ derived from the study of the cross-correlation of the of the 53 GHz and 90 GHz maps for $|b| > 20°$. We have performed a bilinear interpolation in the ($Q_{rms-PS}, \Omega$) plane. The contours are 10 sections at 0.1 intervals of the maximum L-value.

## 4 ANALYSIS OF THE *COBE*-DMR DATA

We used two-year data from the most sensitive radiometers at 53 and 90 GHz. To determine the limits on $Q_{rms-PS}$ and $\Omega$ imposed by the *COBE*-DMR cross-correlation, a grid of Monte Carlo data sets was generated for seven selected $\Omega$ values (0.1, 0.2, 0.3, 0.4, 0.6, 0.8, 1) and $Q_{rms-PS}$ between 7 and 30 $\mu$K in 2 $\mu$K steps. For each simulated (A+B)/2 sky map at 53 and 90 GHz, the cross-correlation was computed and the L statistic calculated. The number of realizations was 3200, a sufficient number as found after testing for the convergence of the relevant quantities.

Given the Harrison-Zeldovich primordial spectrum, we searched for the model ($Q_{rms-PS}, \Omega$) that maximized the likelihood L using Monte Carlo data sets. We find a degeneracy in the plane ($Q_{rms-PS}, \Omega$) that maximizes the value of L. To obtain such a relation between $Q_{rms-PS}$ and $\Omega$, an error to the value of $Q_{rms-PS}$ with maximum L for fixed $\Omega$ is assigned according to the spread of $Q_{rms-PS}$ values obtained from the likelihood method with each realization from that $Q_{rms-PS}$ model taken as the input data. We then fit a cubic polynomial to the pairs ($Q_{rms-PS}, \Omega$) with maximum L considering the corresponding error bars. A qualitative origin of such a relation can be understood in terms of the combined contributions of potential fluctuations and the integrated effect due to curvature which results in a partial cancellation near $\Omega \approx 0.4$. The relationship between our estimates of $\Omega$ and $Q_{rms-PS}$ can be approximated by a cubic polynomial: $Q_{rms-PS}(\Omega) = 10.67 + 55.81\Omega - 128.59\Omega^2 + 81.26\Omega^3$ $\mu$K. The maximum L is obtained for $\Omega = 0.1, Q_{rms-PS} = 15$ $\mu$K, although the likelihood is not significantly worse at higher values of $\Omega$ or near 1. In fact, all ($\Omega, Q_{rms-PS}$) values satisfying the previous cubic relationship are statistically similar. Figure 2 shows L versus $\Omega$ and $Q_{rms-PS}$ and also 10 contours at levels 0.1 of the maximum L-value. Assuming a Bayesian interpretation of L, one can estimate the 2D regions of confidence at different levels ($1\sigma, 2\sigma$, and $3\sigma$) as seen in Fig. 3 where it is clear that above $2\sigma$ all $\Omega$ models are statistically equivalent. Thus, *COBE*-DMR two-year data do not favour a flat universe with $\Omega = 1$. Taking into account an analogous result by Gorski et al (1995) for a different primordial power spectrum (the Ratra-Peebles one), in the sense that the *COBE*-DMR two-year data are not able to provide any specific answer about $\Omega$, we can also conclude that the mentioned result is not specific of the HZ power spectrum but can be considered as a general result independently of the primordial power spectrum assumed.

**Figure 3.** Confidence regions at $1\sigma, 2\sigma$ and $3\sigma$ obtained from the likelihood L assuming a Bayesian interpretation. Above the $2\sigma$ level, there is no any favoured value of $\Omega$.

In table 1 we give the relationship between the normalization amplitude $A$ of the primordial power spectrum and the values of the parameter $Q_{rms-PS}$ obtained in our maximum likelihood analysis. The corresponding rms mass fluctuation at $8h^{-1}$ Mpc, $\sigma_8$, for the different $\Omega$ models considered here is derived for several pairs of values of $\Omega_b$, $h$. It is clear that all models with $\Omega < 0.3$ are inconsistent with observations of clustering at small scales.

**Table 1.** Normalization of the primordial power spectrum $A$ for different values of $\Omega$. $A(20\mu K)$ is the amplitude in $P(k) = Ak$ that gives an *rms* quadrupole amplitude of $20\mu K$, while $A(DMR)$ refers to the COBE-DMR quadrupole normalization. $A$ is given in units $h^{-4}$ Mpc$^4$. The resulting $\sigma_8(h, \Omega_b)$ is also given. $Q_{rms-PS}$ is the value favoured by our analysis for the different values of $\Omega$.

## 5  CONCLUSIONS

Based on the standard likelihood statistic, we have considered the test for the temperature cross-correlation of sky maps and we are able to constrain the quadrupole normalization amplitude $Q_{rms-PS}$ and the density parameter $\Omega$ of the density fluctuation power spectrum at recombination with the *COBE*-DMR two-year data. It is found that the two parameters lie within a region around the polynomial $Q_{rms-PS}(\Omega) = 10.67 + 55.81\Omega - 128.59\Omega^2 + 81.26\Omega^3$ $\mu K$ for a Harrison-Zeldovich spectrum .

We have also studied the effect of cosmic plus sampling variance on the determination of the density parameter $\Omega$ when using the cross-correlation as the statistical quantity in the comparison with the *COBE*-DMR data. The result depends on the 'true' $\Omega$ model assumed but generically a bimodal distribution appears, i.e. there exists another $\Omega$-model which is indistinguishable from the 'true' one.

We can also conclude that the scale invariant spectrum predicted by inflation and the flatness of the universe are neither ruled out nor directly implied by the *COBE*-DMR two-year data.

## ACKNOWLEDGEMENTS

LC, EMG and JLS were supported in part by the European Union, Human Capital and Mobility programme of the European Union, contract number ERBCHRXCT920079, and the Spanish DGICYT, project number PB92-0434-C02-02; ST by the European Union contract CI1-CT92-0013; and LC by a Fulbright fellowship. The *COBE*-DMR datasets, developed by NASA Goddard Space flight Center under the guidance of the *COBE* Science Working Group, were provided by the NSSDC. NS acknowledes support from Center for Particle Astrophysics at University of California, Berkeley.

Table 1: Normalization of the primordial power spectrum $A$ for different values of $\Omega$. $A(20\mu K)$ is the amplitude in $P(k) = Ak$ that gives an *rms* quadrupole amplitude of $20\mu K$, while $A(DMR)$ refers to the COBE-DMR quadrupole normalization. $A$ is given in units $h^{-4}$ Mpc$^4$. The resulting $\sigma_8(h, \Omega_b)$ is also given. $Q_{rms-PS}$ is the value favoured by our analysis for the different values of $\Omega$.

| $\Omega_0$ | $A(20\mu K)$ ($\times 10^6$) | $A$(DMR) ($\times 10^6$) | $\sigma_8(0.5, 0.03)$ | $\sigma_8(0.5, 0.06)$ | $\sigma_8(0.8, 0.03)$ | $\sigma_8(0.8, 0.06)$ | $Q_{rms-PS}$ ($\mu K$) |
|---|---|---|---|---|---|---|---|
| 0.1 | 2.894 | 1.638 | 0.0740 | 0.04245 | 0.1283 | 0.06493 | 15.05 |
| 0.2 | 1.590 | 1.195 | 0.2037 | 0.1578 | 0.3580 | 0.2641 | 17.34 |
| 0.3 | 1.334 | 1.084 | 0.3566 | 0.3011 | 0.6185 | 0.5061 | 18.03 |
| 0.4 | 1.353 | 1.050 | 0.5250 | 0.4620 | 0.8952 | 0.7703 | 17.62 |
| 0.6 | 1.894 | 1.125 | 0.9197 | 0.8433 | 0.1520 | 1.375 | 15.42 |
| 0.8 | 2.179 | 1.165 | 1.318 | 1.233 | 0.2121 | 1.965 | 14.63 |
| 1.0 | 0.798 | 0.732 | 1.337 | 1.268 | 0.2086 | 1.965 | 19.15 |